# Automatic Mapping of the Indoor World with Personal Radars

**Anna Guerra, University of Bologna, Francesco Guidi, University of Bologna, Gianni Pasolini, University of Bologna, Antonio Clemente, CEA LETI, Raffaele D'Errico, CEA LETI, Davide Dardari, University of Bologna**

*Digital maps will revolutionize our experience of perceiving and navigating indoor environments. While today we rely only on the representation of the outdoors, the mapping of indoors is mainly a part of the traditional SLAM problem where robots discover the surrounding and perform self-localization. Nonetheless, robot deployment prevents from a large diffusion and fast mapping of indoors and, further, they are usually equipped with laser and vision technology that fail in scarce visibility conditions. To this end, a possible solution is to turn future personal devices into personal radars as a milestone towards the automatic generation of indoor maps using massive array technology at millimeter-waves, already in place for communications. In this application-oriented paper, we will describe the main achievements attained so far to develop the personal radar concept, using ad-hoc collected experimental data, and by discussing possible future directions of investigation.*

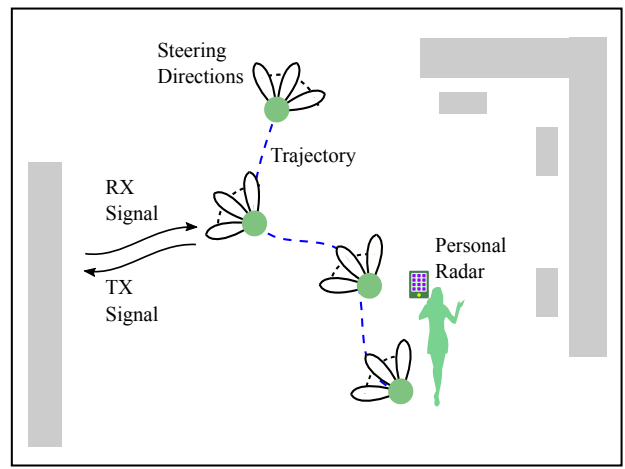

**Figure 1** Schematic example of the personal radar principle in indoor environments.

## Introduction

Smartphones and wearable devices are becoming an outstanding mean for the automatic sensing of environmental parameters thanks to their pervasive diffusion and the increasing number of embedded sensors. In the next future, they might also be able to derive the map of indoor environments, thanks to additional proactive sensing capabilities that will turn them into personal radars [1] [2]. In particular, in evolutionary 5G and beyond scenarios, it is expected that massive arrays working in the mm-wave bands will be integrated on personal devices for enhancing the communication performance, and, thus, the same hardware may be also exploited to implement radar functionalities [2]. Consequently, personal devices will be able to accurately scan the environment by transmitting probe radio signals via the generation of narrow radio beams pointing in different directions (beamsteering), and by receiving the signal reflected by the surrounding [2]. Processing such echoes will allow to retrieve ranging and bearing information, and will make it possible to derive the maps of indoor spaces, with no need of any dedicated infrastructure.

Nowadays, when no infrastructure is present, traditional SLAM methods are usually adopted, where self-localization and mapping are performed using laser-based radars (lidars) [3]. Unfortunately, lidars are usually expensive, energy-hungry, and their usage is constrained for safety reasons. Other solutions entail the adoption of camera-based devices (e.g., V-SLAM) to discriminate scene features, which however become ineffective when the distance between the camera and the captured scene is too large [4]. Moreover, both approaches (either based on cameras or lidars) need perfect visibility conditions and a mechanical steering of the scanners. When moving towards a large-scale diffusion, these systems do not represent a feasible solution to be embedded in personal devices, which are expected to discover the environment in an automatic fashion, i.e., without any user supervision [2]. To this end, compared to other solutions, the radio technology might be more advantageous, relying on the same hardware (the RF front-end) used for communications. Currently, radars at 77 GHz have been investigated for target tracking applications, without considering the mapping problem [5]. Thus, the peculiarities of mm-wave massive antenna arrays still require ad-hoc approaches both for the design of the mapping algorithms [3].

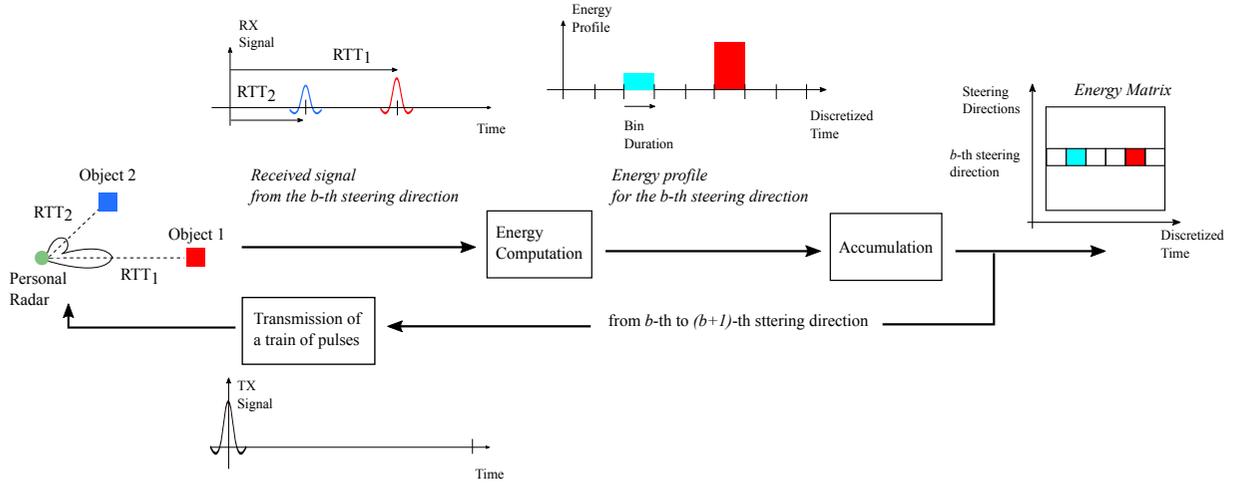

*Figure 2 Personal radar loop for energy matrix reconstruction.*

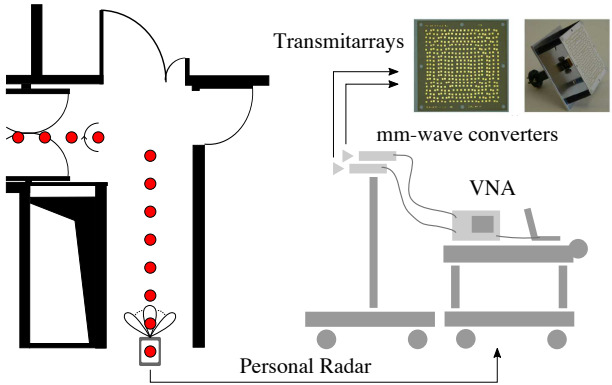

*Figure 3. Measurement scenario (left), where a personal radar follows the trajectory indicated by red dots, and measurement setup (right).*

Within this context, we believe that the time is mature for an overview of all the diverse aspects that must be addressed to turn the personal radar concept into reality.

This application-oriented article introduces the main steps conceived so far to use mm-wave massive arrays for indoor automatic mapping, discussing antenna, propagation and algorithmic aspects under a unique framework. Given the application-oriented scope of this manuscript, the discussion of the main theoretical findings is accompanied by simulative or experimental validation.

## The Personal Radar Idea

The automatic mapping concept relies on the idea that devices, equipped with personal radars, automatically scan the environment and derive the correspondent maps. Differently from laser-based systems, a massive array is used to generate a narrow radio-beam that can be electronically steered in different directions, without the need of an active role of the device owner.

For each of these steering directions, the transmitter sends a train of pulses, and the co-located receiver collects the corresponding backscattered signal (that undergoes the two-way channel response). In Fig. 1, the dashed line indicates the radar trajectory, and the solid lines the signal exchanges.

Once the backscattered measurements have been collected, a low complexity energy-based approach can be adopted to process the received signals, following the steps of Fig. 2:

1. *Energy computation*: For each steering direction, the received energy is computed by integrating the square of the received signal echo in temporal bin of duration approximately equal to signal bandwidth inverse;
2. *Accumulation*: In order to increase the signal-to-noise ratio (SNR), the computed received energy is accumulated over the number of transmitted pulses;
3. *Energy matrix construction*: This procedure is repeated for each steering direction, thus giving a temporal-angular energy matrix which is the input for the channel modelling and the mapping algorithms.

Unfortunately, even with large antenna arrays, the adopted antenna radiation patterns are far from being laser-like, and they usually present side-lobes that might create ambiguities in the mapping procedure. In this regard, Fig. 2 shows an example of what might happen for a fixed steering direction, given the presence of objects in the main and in side-lobe direction, respectively. In the depicted example, the reconstructed energy profile wrongly associates a contribution to the considered steering direction due to the signal arriving from the side lobe with a delay indicated as $RTT_2$. While the problem has been largely investigated and solved for coherent receivers [6] or it is not present in laser-based systems (e.g., lidar), the impact of side-lobes on the mapping performance should be accounted for in personal radars.

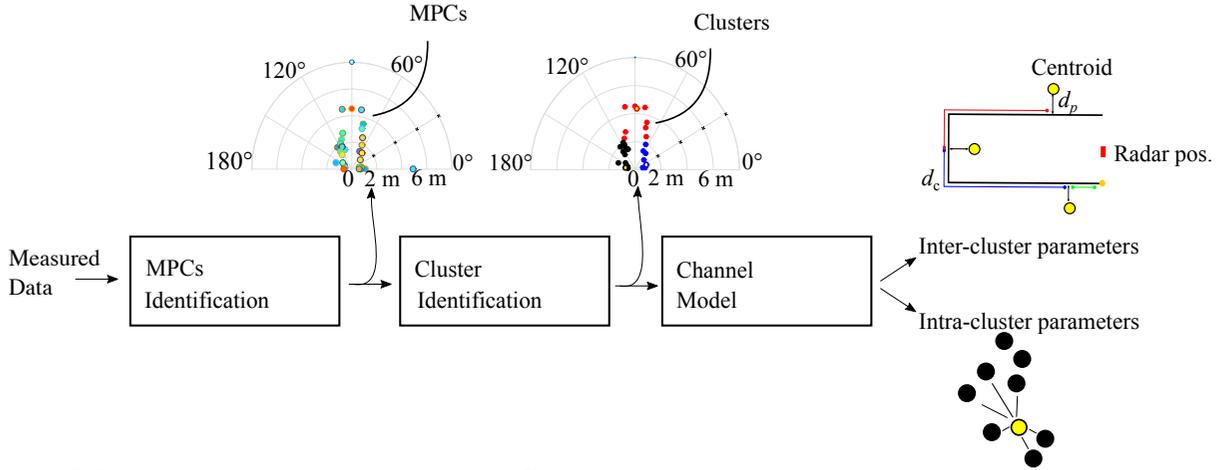

*Figure 4 Followed steps for radar channel modelling in [10].*

## The Measurement Campaign

A measurement campaign was carried out to gather real world backscattered energy profiles. As for massive arrays, two linearly polarized transmitarrays (TAs) were adopted, each with 400 antennas and an overall dimension of $5 \times 5$ cm$^2$ [7]. The two TAs were placed in a quasi-monostatic configuration in order to isolate the transmitting and receiving channels. Measurements were collected in the $57.5 - 63.5$ GHz band. Since the adopted non-reconfigurable TAs [7] do not allow to electronically steer the antenna beam, the antennas were placed on a mechanical support whose position and orientation varied at each time instant to emulate the personal radar movements and the beamsteering (see Fig. 3-right).

The radar was moved in 12 positions in a corridor whose layout is reported in Fig. 3-left, with spatial points separated of 0.405 m. Then, for each tested position, the radar was rotated in the semi-plane with a step of 5°.

This set-up allowed to collect the channel responses for different spatial positions and angular steering directions. Starting from these data, it has been possible to characterize the radar channel and to emulate its behaviour through simulations.

## The Backscattering Channel Model

This section summarizes the steps needed to characterize the channel experienced by personal radars. In this regard, it is worth stressing that the availability of a statistical model of the backscattering channel is fundamental both for the design of the mapping algorithm and for the assessment of the personal radar performance. Further, this investigation fills a gap in the literature, where the majority of works on mm-wave channel modeling has focused on one-way communications, as in [8]. One of the few examples of two-way channel characterization can be found in [9], but it dealt with ultrawide bandwidth radio frequency identification. Here instead the focus is on the clutter channel response for personal radar mapping at mm-waves starting from the measurements collected during the previously described campaign.

The flowchart of the main steps is shown in Fig. 4. In particular, the first task is the identification of the multipath components (MPCs). To this purpose, starting from the energy matrix created for all steering directions and temporal bins, the estimation of MPCs aims at inferring the number, power, delay and direction of MPCs by discerning them from components coming from side-lobes or due to noise.

*Table 1 Channel Parameters for the corridor scenario of Fig. 3.*

| Parameter | Standard Deviation | Distribution |
|---|---|---|
| Inter-cluster arrival rate | 0.21 [1/ns] | Exponential |
| Inter-cluster perimeter distance | 5.02 [ns] | - |
| Intra-cluster arrival rate | 0.43 [1/ns] | Exponential |
| Intra-cluster AOA | 16.9 (°) | Laplacian |

To this end, it is important to jointly operate both in the delay and angular domains by setting filtering windows based on the transmitted pulse duration and on the antenna pattern [10]. This avoids a wrong association of the selected MPCs with side-lobe directions. Fig. 4 shows, for a given radar position, an example of estimated MPCs where colors represent their power [10].

At the end of this phase, it is usually observed that MPCs are grouped in clusters, i.e., groups of components with similar delays and angular directions. Consequently, a *K*-means algorithm was used to identify the number of clusters and their centroids and to classify MPCs within clusters [10]. Fig. 4 shows an example of how paths are grouped into clusters, with MPCs belonging to the same cluster are represented with the same color. After these two preliminary phases, all parameters (number, delays and angles of MPCs and clusters) are available to be used for modeling the channel.

To tackle this objective, a modified Saleh-Valenzuela channel model can be adopted to describe the statistics of the intra- and inter-cluster parameters. In a completely innovative manner, the cluster centroid distribution is characterized as a function of the environment layout. In fact, by considering the map perimeter, two figures of merits are of interest: (i) the statistical description of the cluster centroid distance from the closest edge (indicated with $d_p$ in Fig. 4); and (ii) the distance along the perimeter of the cluster centroid projections (indicated with $d_c$ in Fig. 4).

Successively, the intra-cluster parameters are analyzed in a more classical way, by modelling the intra-cluster angle-of-arrivals and time-of-arrivals.

A more detailed description of the steps for channel characterization can be found in [10] whose results are summarized in Table I.

## Automatic Mapping Techniques

We now briefly review some mapping solutions proposed for personal radar applications. In the mapping literature, two types of models are generally privileged: feature- and grid-based approaches [11], [12], [13]. The first ones permit to describe the environment with landmarks that can be easily discerned thanks to accurate sensing technologies (e.g., laser). However, since the radiation pattern of massive arrays is less sharp than laser beams, grid-based approaches are preferable for radar applications [2]. According to such approaches, the environment is represented with a grid of elementary cells, characterized by the value of a given property (e.g., the root radar cross section (RRCS)), that form the state to be estimated by a mapping procedure.

Among stochastic approaches exploiting the grid discretization of the environment, methods based on Kalman filtering or on occupancy grid (OG) offer a good trade-off between performance and complexity with respect to particle filtering methods [11].

### Extended Kalman Filter Approach

In Bayesian filtering approach, the *a-posteriori* probability function (belief), i.e. $b(\mathbf{m}^{(k)}) = f(\mathbf{m}^{(k)} | \mathbf{z}^{(1:k)})$, of the state $\mathbf{m}^{(k)}$ given the history of measurements, i.e. $\mathbf{z}^{(1:k)}$, until the current time instant $k$, is estimated. In our mapping scenario, the state contains the RRCS values of each cell of the grid-map, whereas the measurements are the received energy profiles for each steering direction as preciously described. The underlying assumption is that the RRCS of each cell can be approximated with a Gaussian random variable for mathematical tractability, thus allowing the adoption of the Extended Kalman filter (EKF). Under this assumption, the a-posteriori probability is completely characterized by its mean vector (used as the final estimate of the map) and its covariance matrix (used as the confidence on the estimates).

The way in which the EKF-based approach can be applied to the personal radar problem has been fully described in [2] starting from the definition of a proper Gaussian observation model, which is a non-linear radar range equation, and the assessment of its Gaussianity (valid for a large number of transmitting signals). Even if EKF-based approaches are widely exploited for estimating the mobile position and orientation together with the map of the environment, the mapping approach proposed in [2] differentiates from the state-of-the-art for two main reasons: (i) despite it makes use of a grid representation usually suitable for laser-beams, it is able to account for the array pattern and for the consequent "illumination" of multiple cells in the observation model; (ii) it considers all the available raw measurements ("soft" decision) as an input for the mapping process instead of relying on a classic two-step approach where, for each test direction, the obstacle is detected and its distance/angle measured.

### Occupancy Grid Approach

The drawback of EKF-approaches is that the bi-modal nature of the cell (i.e., a cell can be either empty or occupied) is completely neglected. The easiest way to cope with this issue is to adopt an OG algorithm.

The OG map is a probabilistic map in which each cell or pixel is characterized by a certain probability of being occupied [14]. The OG approach takes its roots from the Bayes formula to infer the belief, i.e. the *a-posteriori* probability over the occupancy of each cell of the map. In fact, as before, the objective is to estimate the belief of the map $\mathbf{m}^{(k)}$, where each cell $m_i^{(k)} = \{0,1\}$ is a binary random variable (RV) represented by the probability of its occupancy, given the history of measurements.

Considering the whole map, the mapping problem becomes a maximum a posteriori estimation in a high-dimensional space and, thus, its direct computation is prohibitive. In order to reduce the complexity, OG assumes the independence among the beliefs of the cells and in the observation model, instead of computing the joint conditional probability density function in its native form. While this is reasonable for laser observations, it becomes quite inaccurate for radar measurements due to the presence of side lobes that may generate ambiguities [14]. Therefore, in [15] a low-complexity method accounting for the correlation among cells due to antenna pattern is proposed. Specifically, the beliefs are expressed not only as a function of the cell pointed by the radar, but also of all the others intercepted and contributing to the same energy bin according to the radiation pattern, as shown in Fig. 2.

## Case Study

We now present some mapping examples obtained through the adoption of the above introduced algorithms. Both simulated and measured data were used to assess the personal radar performance.

### Measurement-based and Model-based Maps

In this section, we compare results obtained by using measurement and model-based data. To this purpose, we exploit the EKF-based mapping carried out using simulated observations as in [2] or real-world measurement data as in [10]. The grid cells have an area of 0.01 m$^2$ each and the radar moves along a path, denoted by a dashed line in Fig. 5 [10]. At a first instance, we are interested in solving only the mapping problem, and, hence, the radar position and orientation are assumed known. A complete list of the adopted parameters is reported in [2] as well as the situation in which there are uncertainties on radar locations.

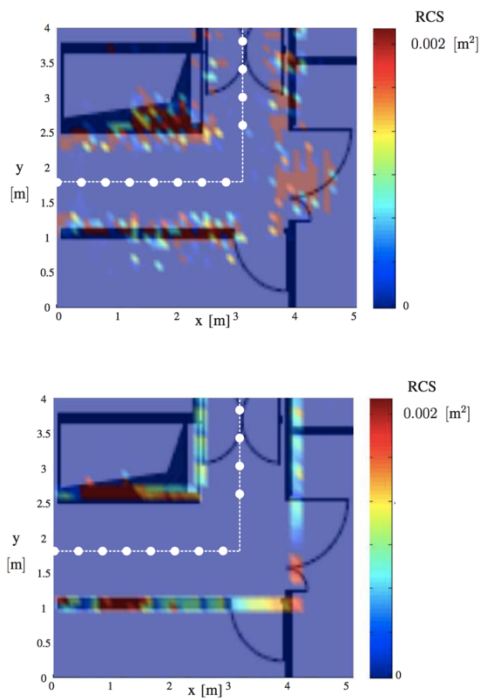

*Figure 5* Mapping results using 1-bit 400-antenna arrays exploiting measured (top) and simulated (bottom) data.

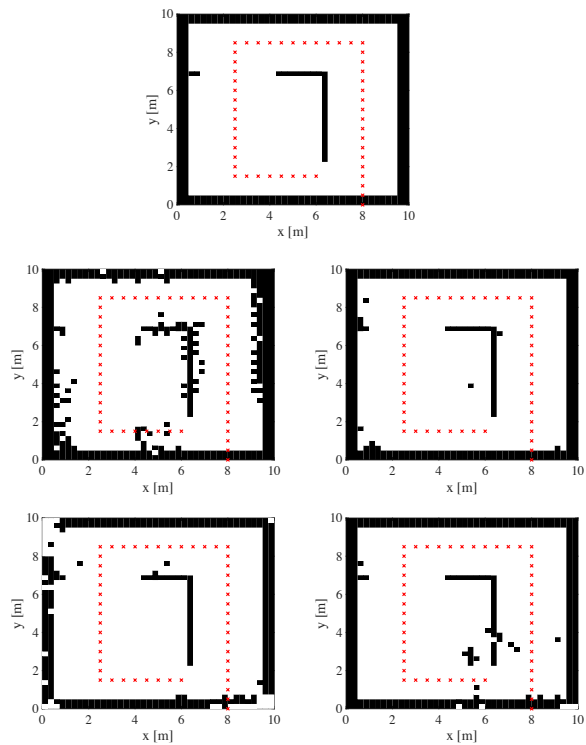

*Figure 6* EKF-OG comparison. Top: reference map; Middle: Occupancy Grid; Bottom: EKF; Left: 16 array antennas; Right: 100 array antennas.

Fig. 5 shows the mapping results obtained using measured channel responses (top) and simulated ones (bottom). Notably, in both cases there is a good agreement between the real map and its estimation. The discrepancies observed in Fig. 5-top can be ascribed to the mismatch between the measurements and the model included in the EKF algorithm, which is simplistic in terms of wave propagation, and RCS characteristics [2]. It is worth to observe that the results have been obtained considering only one user. However, the personal radar concept can be extended to multiple users moving in the same area and performing data fusion, thus enhancing the mapping performance thanks to the increased amount and diversity of the available information.

### EKF- and OG-based Maps

When comparing the performance of the EKF- and OG-based mapping algorithms, simulated channel responses where used as they are less noisy and facilitate the performance comparison. Furthermore, bitmap pictures have been obtained by applying a threshold $\eta$ at the end of the mapping process to decide whether a cell is occupied (i.e., $b(m_i) > \eta$) or not (i.e., $b(m_i) < \eta$).

In the maps of Fig. 6, the color of each cell represents the cell occupancy: a white cell is free, and a black one is occupied. A detailed description of the simulation set-up is reported in [15]. Since the prior initialization significantly impacts the performance, maps for each method are provided in their best conditions. For example, in the OG each cell is initialized to 0.5 (i.e., complete uncertainty), whereas the EKF case is initialized to an empty map. The threshold for deciding whether a cell is empty or not is chosen according to the considerations in [2].

From Fig. 6 it emerges the impact of the number $N$ of antennas on the map reconstruction: as expected, with $N = 16$ (left), the number of pixels with a wrong occupancy value is higher with respect to $N = 100$ (right).

On the other side, the OG and the EKF approaches perform similarly for map reconstruction. Nonetheless, it is worth to mention that results greatly depend on the chosen initialization parameters, especially for the EKF approach. Indeed, by varying the aforementioned cell initialization of 40%, for $N = 16$ the ISI increases of about 5% for the OG and of 220% for the EKF approach, meaning that the latter less likely converges to a satisfactory solution. Thus, the OG approach results to be more robust even though it cannot provide, as a side result, an electromagnetic description of each cell (e.g., RCS value). Whenever the problem becomes the estimation of physical quantities, related for instance to materials, EKF is to be privileged because of its capability to infer the mean value of the parameter of interest and its confidence level.

## Conclusions & Outlook

In this paper, the personal radar application for the automatic mapping of the indoor world has been described, with a system overview spanning from backscattering radar channel model and ad-hoc signal processing algorithms for environment mapping. Results, which have been obtained using also experimental data, are promising and leave the door open for putting in practice this application. On the other side, there is still the lack of a dedicated study on the expected energy consumption, in order to find a trade-off between the achievable performance and an affordable architecture complexity for a practical implementation on mobiles.

In addition, future crowd-sensing techniques can be explored to boost the mapping performance by combining observations or estimates of different users. Maps might be shared among all the users within a certain area for navigation assistance in bad visibility conditions (due to smoke, for instance) or for supporting visually impaired persons. Vision information coming from smartphone-embedded cameras may be also fused with radar-based information to obtain more reliable 3D descriptions of the physical space. In this sense, personal radars could be also conceived as an add-on for current vision-based SLAM systems, merging the advantages of both radio and visible light frequencies.

## Acknowledgment

This work has received funding from the European Union's Horizon 2020 research and innovation programme under the Marie Sklodowska-Curie project AirSens (grant no. 793581)